\documentstyle[preprint,prl,aps,epsf,epsfig,psfig,euscript,graphics,fancybox,fancyheadings,rotating, 
		times,subfigure,layout,hangcaption]{revtex}

\begin{document}
\draft


\def\isn{$^2$}
\def\sphn{$^1$}
\def\md{$^4$}
\def\jlab{$^6$}
\def\mit{$^5$}
\def\rutgers{$^7$}
\def\bale{$^3$}
\def\lns{$^{10}$}
\def\ncat{$^{9}$}
\def\fiu{$^{8}$}
\def\ipn{$^{12}$}
\def\yerevan{$^{11}$}

\title{Phenomenology of the deuteron electromagnetic form factors}

\author{D.~Abbott,\jlab\   
	A.~Ahmidouch,\mit$^,$\ncat\  
	H.~Anklin,\fiu\   
	J.~Arvieux,\lns$^,$\ipn\   
	J.~Ball,\sphn$^,$\lns\   
	S.~Beedoe,\ncat\   
	E.J.~Beise,\md\   
	L.~Bimbot,\ipn\    
	W.~Boeglin,\fiu\   
	H.~Breuer,\md\   
	R.~Carlini,\jlab\    
	N.S.~Chant,\md\   
	S.~Danagoulian,\jlab$^,$\ncat\   
	K.~Dow,\mit\   
	J.-E.~Ducret,\sphn\    
	J.~Dunne,\jlab\    
	L.~Ewell,\md\   
	L.~Eyraud,\isn\   
	C.~Furget,\isn\   
	M.~Gar\c con,\sphn\    
	R.~Gilman,\jlab$^,$\rutgers\   
	C.~Glashausser,\rutgers\   
	P.~Gueye,\jlab\   
	K.~Gustafsson,\md\   
	K.~Hafidi,\sphn\    
	A.~Honegger,\bale\    
	J.~Jourdan,\bale\   
	S.~Kox,\isn\   
	G.~Kumbartzki,\rutgers\   
	L.~Lu,\isn\    
	A.~Lung,\md\      
	P.~Markowitz,\fiu\   
	J.~McIntyre,\rutgers\   
	D.~Meekins,\jlab\   
	F.~Merchez,\isn\    
	J.~Mitchell,\jlab\
	R.~Mohring,\md\    
	S.~Mtingwa,\ncat\   
	H.~Mrktchyan,\yerevan\   
	D.~Pitz,\sphn$^,$\md$^,$\jlab\    
	L.~Qin,\jlab\    
	R.~Ransome,\rutgers\   
	J.-S.~R\'eal,\isn\   
	P.G.~Roos,\md\    
	P.~Rutt,\rutgers\    
	R.~Sawafta,\ncat\   
	S.~Stepanyan,\yerevan\      
	R.~Tieulent,\isn\   
	E.~Tomasi-Gustafsson,\sphn$^,$\lns\    
	W.~Turchinetz,\mit\   
	K.~Vansyoc,\jlab\footnote{On leave from Old Dominion University, USA}\  
	J.~Volmer,\jlab\footnote{On leave from Vrije Universiteit, Amsterdam, Netherlands}\   
	E.~Voutier,\isn\      
	C.~Williamson,\mit\   	
	S.A.~Wood,\jlab\    
	C.~Yan,\jlab\   
	J.~Zhao,\bale\   
	and 
	W.~Zhao\mit\\
        (The Jefferson Lab t$_{\hbox{20}}$ collaboration)
       }

\address{
\sphn DAPNIA/SPhN, CEA/Saclay, 91191 Gif-sur-Yvette, France\\ 
\isn ISN, IN2P3-UJF, 38026 Grenoble, France\\
\bale Dept. of Physics, University of Basel, Switzerland\\
\md University of Maryland, College Park, MD 20742, USA\\
\mit M.I.T.-Bates Linear Accelerator, Middleton, MA 01949, USA\\
\jlab Thomas Jefferson National Accelerator Facility, Newport News, VA 23606, USA\\
\rutgers Rutgers University, Piscataway, NJ 08855, USA\\
\fiu Florida International University, Miami, FL 33199, USA\\
\ncat North Carolina A. \& T. State University, Greensboro, NC 27411, USA\\
\lns LNS-Saclay,  91191 Gif-sur-Yvette, France\\
\yerevan Yerevan Physics Institute, 375036 Yerevan, Armenia\\
\ipn IPNO, IN2P3, BP 1, 91406 Orsay, France\\
}

\date{Re-submitted to EPJ A, February 25, 2000}

\maketitle

\bigskip
\bigskip\bigskip
\bigskip
\begin{abstract}
A rigorous extraction of the deuteron charge form factors
from tensor polarization data in elastic electron-deuteron
scattering, at given values of 
the 4-momentum transfer, is presented.
Then the world data for elastic electron-deuteron scattering is used
to parameterize, in three different ways, the three electromagnetic
form factors of the deuteron in the 4-momentum transfer range 
0-7 fm$^{-1}$. This procedure is made possible
with the advent of recent polarization measurements. 
The parameterizations allow a phenomenological characterization of the
deuteron electromagnetic structure. They
can be used to remove ambiguities in the form factors extraction
from future polarization data.
\end{abstract}

\narrowtext

{\bf 1 Introduction}

The deuteron, as the only two-nucleon bound state, has been the
subject of many theoretical and experimental investigations. Since it
has spin 1, its electromagnetic structure is described by three
form factors, charge monopole $G_C$, charge quadrupole $G_Q $
and magnetic dipole $G_M$, assuming P- and T-invariance. 
Measurements of elastic electron
deuteron scattering observables  provide quadratic combinations
of these form factors. Since most of the data available come from
differential cross section measurements, it has been customary, both in
the data presentation and in the comparison with theoretical models, 
to use the two structure functions $A$ and $B$ defined hereafter,
extracted from the cross section data by a Rosenbluth separation \cite{ROS50}.
With the advent of tensor polarimeters and tensor polarized internal
targets, polarization observables have been measured as well, which allow
the separation of the two charge form factors. 

The purpose of this work is twofold.
First, in Sect. 2,
the calculation of $G_C$ and $G_Q$,
at given values of the 4-momentum transfer $Q$, 
from polarization data together with (interpolated) $A$ and $B$ data
is reexamined and updated with respect to previous work.

Then, in Sect. 3, parameterizations of the three
deuteron form factors, in the 4-momentum transfer range $Q=0-7$~fm$^{-1}$,
are provided.
Above 7 fm$^{-1}$, only small angle cross section data are available,
preventing the separate determination of the three form factors.
We have determined the three deuteron electromagnetic form factors
by fitting directly
the measured differential cross 
section~\cite{ABB99,AKI79,ALE99,ARN75,AUF85,BEN66,BOS90,BUC65,CRA85,DRI62,ELI69,FRI60,GAL71,GAN72,GRO66,MAR77,PLA90,RAN67,SIM81}
and polarization~\cite{ABB00,BOD91,BOU99,DMI85,FER96,GAR94,GIL90,SCH84,VOI86}
observables. 
This procedure eliminates the need for an intermediate determination 
of $A$ and $B$, and results in a more realistic evaluation of errors 
for the form factors.

One parameterization is used for a determination of the node of the
charge form factor $G_C$, while the application of the
work of Ref.\cite{KOB95} allows the determination of reduced form factors
in a helicity basis. The accuracy in the determination
of these form factors is limited
by the assumption of a one-photon exchange mechanism in the first
order Born approximation at low $Q$, and by the accuracy of the data at 
intermediate to high momentum transfers. A third parameterization
was recently applied for a precise determination
of the rms--charge radius of the deuteron 
\cite{SIC98}. At low $Q$, Coulomb distortion was
taken into account to extract precise values of $G_C$. 
Applying this correction
resolved an old discrepancy between
the deuteron radius determined via $(e,e')$ and 
N--N scattering\cite{KLA86}. In the intermediate to high 
$Q$-range, other corrections such as the
double scattering contribution to two photon exchange~\cite{2PH} 
should be considered, but they are at present neither accurately calculated
nor experimentally determined.

\vspace{5mm}
{\bf 2 Observables and form factors}

{\bf 2.1 e-d observables}

Assuming single photon exchange, the electron-deuteron unpolarized 
elastic differential cross section can be written as
\begin{equation}
\frac{d\sigma}{d\Omega} = 
\sigma_{NS} \cdot  
\left[ G^2_C(Q^2) + \frac{8}{9} \eta^2 G^2_Q(Q^2) 
	    + \frac{2}{3} \eta \varepsilon^{-1}(Q^2,\theta_e) G^2_M(Q^2)
\right]
\equiv \sigma_{NS} \cdot S,
\label{eq:sigma}
\end{equation}
where 
$\sigma_{NS}$ 
is the  Mott differential cross section
multiplied by the deuteron recoil factor,    
$\theta_e$ the  electron scattering angle,
$\eta = Q^2/4M_d^2$, $M_d$ the deuteron mass;
$\varepsilon =  [1 + 2(1+\eta)\tan^2(\theta_e/2)]^{-1} $
is related to the virtual photon polarization.
The quantity $S \equiv A + B \tan^2(\theta_e/2)$ 
defines the usual $A$ and $B$ elastic structure functions. 

The tensor polarization observables $t_{2q}$, or equivalently the
analyzing powers $T_{2q}$, have
been measured as well. Their expression as a function of the three
form factors, still in the one-photon exchange approximation, is
given by:
\begin{eqnarray}
-\sqrt{2}\cdot S
\cdot t_{20} & = &
\frac{8}{3} \eta G_C G_Q + \frac{8}{9} \eta^2 G_Q^2
+\frac{1}{3} \eta \varepsilon^{-1} G_M^2		\label{eq:t20}
\\
\sqrt{3}\cdot S
\cdot t_{21} &  = & 
2\eta\left(\eta + \eta^2\sin^2\frac{\theta_e}{2}\right)^{1/2} 
G_M G_Q \sec\frac{\theta_e}{2}				\label{eq:t21}
\\
-2\sqrt{3}\cdot S
\cdot t_{22}  & =  & \eta G_M^2.			\label{eq:t22}
\end{eqnarray}

{\bf 2.2 Calculation of $G_C$ and $G_Q$}

The charge form factors are here extracted from
$t_{20}(Q,\theta_e)$ data, together with $A(Q)$ and $B(Q)$ (interpolated) data. 
The analyses presented in \cite{BOD91,GAR94}
need to be updated, because of new $t_{20}$~\cite{ABB00,BOU99,FER96}
and $A$~\cite{ABB99,ALE99} data. In particular, the parameterization of $A$ 
used in \cite{GAR94} gave a very small weight to the then only existing high $Q$
data~\cite{ARN75} and is lower than the new data~\cite{ABB99,ALE99}
around 4.5 fm$^{-1}$. Furthermore,
we present here a more compact solution and a more rigorous treatment 
of errors.

For our purpose, it is useful to define new quantities
$A_0 \equiv A - B/2(1+\eta)$ and
$\tilde{t}_{20}$ \cite{GAR94}, derived respectively from $A$ and $t_{20}$
by eliminating the magnetic contribution:

\begin{equation}
\tilde{t}_{20} \equiv 
			-\frac {\frac{8}{3} \eta G_C G_Q + \frac{8}{9} \eta^2 G_Q^2}
                                                 {\sqrt{2}\ (G^2_C + \frac{8}{9} \eta^2 G^2_Q)}
               =      \frac {S\cdot t_{20} + B/4\sqrt{2}\varepsilon (1+\eta)}{A_0}
\label{eq:t20t}               
\end{equation}
 Using the reduced form factors $g_C = G_C/\sqrt{A_0}$ and 
$g_Q = 2\eta G_Q/3\sqrt{A_0}$, (\ref{eq:sigma},\ref{eq:t20},\ref{eq:t20t}) lead to:
\begin{eqnarray}
g_C^2 + 2 g_Q^2& = &  1\\
2 g_C g_Q +  g_Q^2& = & p \equiv - \tilde{t}_{20}/\sqrt{2} \label{eq:syst_g2}
\end{eqnarray}
where $p$ (or conventionnally $p_{ZZ}$)
is the tensor polarization in Cartesian notation (also called alignment).
There are four solutions to these equations given by
\begin{equation}
(g_Q^{\pm})^2  =   \frac{2+p \pm \sqrt{\Delta}}{9} \label{eq:gq}
\end{equation}
with $\Delta = 8(1-p)(\frac{1}{2}+p)$ and $g_C^{\pm}$ from (\ref{eq:syst_g2}).
The physical solution is easily selected at small $Q$ from
the static moments ($g_C(0)=1,\ g_Q(0)=0$). It corresponds to
the choice of a minus sign in (\ref{eq:gq}) and of $g_Q>0$.
Since $\tilde{t}_{20}$ and $t_{21}$, both proportional to $G_Q$,
do not cross zero at a same value of $Q$ \cite{ABB00,GAR94},
$g_Q$ has to remain positive over the whole range considered in this
work. The two remaining solutions 
($g_Q^+,g_C^+$) and ($g_Q^-,g_C^-$) cross each other 
at values $Q_{min}$ and $Q_{max}$ where $\tilde{t}_{20}$ 
reaches its extrema $-\sqrt{2}$ and $+\sqrt{2}/2$ ($\Delta = 0$).
The physical solution must  switch 
from ``$-$" to ``$+$" at $Q=Q_{min}$ and then
back to ``$-$" at $Q_{max}$
in order to ensure a continuity of the form factor derivatives.
For polarization data close to these extrema,  $Q$ may be
below or above the {\sl a priori} unknown $Q_{min}$ (or $Q_{max}$),
and the choice of solution is ambiguous. $Q_{min}$, from our three global 
fits to the $e-d$ data (see Sect. 3), is determined to be close to 3.3 fm$^{-1}$. 
On the other hand, there
 are not enough polarization data to constrain the value of
$Q_{max}$, so that the above mentioned ambiguity remains around
$Q\simeq 6-8$ fm$^{-1}$. This is the case for the
two points at highest $Q$ in \cite{ABB00}. 

An additional complication arises for five 
polarization data points in Refs.\cite{ABB00,BOD91,BOU99,GAR94,GIL90} which 
lay partially outside the physical
region $-\sqrt{2} \leq \tilde{t}_{20} \leq 1/\sqrt{2}$. This situation is
quite probable for points with finite errors close to a physical
limit \cite{PDG98}.
For the sake of extracting $G_C$ and $G_Q$, 
the interval of 68.3\% confidence level 
$[\tilde{t}_{20}-\Delta\tilde{t}_{20},\tilde{t}_{20}+\Delta\tilde{t}_{20}]$, 
and eventually the most probable value $\tilde{t}_{20}$, are
then modified according to the method presented in \cite{FEL98}. The
resulting confidence interval is entirely within the physical
region ($\Delta \geq 0$). In this particular case, the modified values of $p$ 
are used in (\ref{eq:syst_g2},\ref{eq:gq}) instead of the measured ones. As a result
of this procedure, the errors on the form factors may be asymmetric.

The calculated values of $G_C$ and $G_Q$, corresponding to all measurements
of $t_{20}$, are presented in Table 1 and Fig. 1. The later also
shows results of parameterizations to be discussed in Sect. 3. 
Uncertainties come from the quoted
errors in $t_{20}$, combined quadratically with errors on $A$ and $B$
reflecting the spread of the data (for example, at 5 fm$^{-1}$, 
8.5 and 17 \% respectively).
For the two points of highest
$Q$, the two solutions of (\ref{eq:syst_g2},\ref{eq:gq}) are given. 
The first one is preferred, based on
theoretical guidance and on the parameterizations discussed below. 
Only parameterization I (Sect. 3.1) favors the second solution
for the point at $Q=6.64$ fm$^{-1}$. 
Note that $\tilde{t}_{20}$ need not necessarily reach
its maximum allowed value, in which case the first (``$+$") solution
would prevail from $Q=Q_{min}$ up to the undetermined node of $G_Q$,
or to the second minimum of $\tilde{t}_{20}$, whichever occurs first.

\vspace{5mm}
{\bf 3 Parameterization of the form factors}

The three paramaterizations described below are determined 
through a $\chi^2$ minimization involving 269 cross section data 
points~\cite{ABB99,AKI79,ALE99,ARN75,AUF85,BEN66,BOS90,BUC65,CRA85,DRI62,ELI69,FRI60,GAL71,GAN72,GRO66,MAR77,PLA90,RAN67,SIM81} 
and 39 polarization data 
points~\cite{ABB00,BOD91,BOU99,DMI85,FER96,GAR94,GIL90,SCH84,VOI86}.
In most polarization data, and in some cross section data, the systematic uncertainties
are dominant and may vary from point to point in a given experiment. The
error considered in the $\chi^2$ minimization is then the quadratic sum of
the statistical and systematic uncertainties. The uncertainties on the
parameters are  given by the error matrix.
For  data where
an overall normalization uncertainty may apply, the resulting systematic
uncertainty of the fitted parameters have been evaluated by changing each
individual data set by the quoted error and re-fitting the complete data
set. This last procedure was carried on only with parameterization III (Sect. 3.3).

The $\chi^2$ per degree of freedom ($\chi^2/N_{d.f.}$)
all exceed the value of 1, because
of systematic differences between some data sets, at the limit or
beyond the quoted systematic uncertainties. Among the most recent
experiments, this is  the case for the $A$ measurements
of Refs.~\cite{ABB99,ALE99}, and in a lesser extent for the $t_{20}$
measurements of Refs.~\cite{ABB00,GAR94}. The fits then give an
average representation of the  data, though biased toward experiments with 
a larger number of data points.

{\bf 3.1 Parameterization I}

In the first parameterization (I), each form factor is given by:
\begin{equation}
 G_X(Q^2) = G_X(0) \cdot \left[1-\left({Q \over Q_X^0}\right)^2\right] \cdot  
		\left[ 1+\sum_{i=1}^{5}{{a_X}_i Q^{2i}}\right] ^{-1},
\label{eq:parI}
\end{equation}
with $X = C, Q$ or $M$.
This expression has the advantage of displaying
explicitly the first node $Q_X^0$ of each form factor.
The normalizing factors $G_X(0)$ are fixed by the deuteron static moments. 
With 18 free parameters, a fit is obtained with $\chi^2/N_{d.f.}=1.5$.

{\bf 3.2 Parameterization II}

Another parameterization (II) has been proposed by
Kobushkin and Syamtomov \cite{KOB95}.  
Each form factor is
proportional to the square of a dipole nucleon form factor $G_D$
and to a linear combination of reduced helicity transition
amplitudes $g_0, g_1, g_2$:
\begin{equation}
	\left( \begin{array}{c} G_C \\ G_Q \\ G_M \end{array} \right)
	= G_D^2 \left({Q^2 \over 4}\right) \cdot \mathcal{M}(\eta)
	\left( \begin{array}{c} g_0 \\ g_1 \\ g_2 \end{array} \right).
	\label{eq:kob1}
\end{equation}
	Each of these amplitudes is parameterized as a sum
of four Lorentzian factors:
\begin{equation}
	g_k = Q^k \sum_{i=1}^{4} \frac{a_{ki}}{\alpha_{ki}^2 + Q^2}.
	\label{eq:kob2}
\end{equation}
For each $k$, the $\alpha_{ki}^2$ follow an arithmetical suite
defined by 2 independent parameters.
In addition,
an asymptotic behavior
dictated by quark counting rules and helicity rules valid in perturbative
quantum chromodynamics (pQCD), 
together with the normalization conditions at $Q=0$, 
imply 6 relations
between the parameters $a_{ki}$ and $\alpha_{ki}$~\cite{KOB95}.
As a result, each amplitude is described by 4 independent parameters.
New parameters are obtained
here, due on one hand to a newer data base, and 
on the other hand to the fitting of
the differential cross sections instead of $A$ and $B$.
With 12 free parameters, a fit to the data set is obtained
with $\chi^2/N_{d.f.} = 1.8 $,
whereas the original values of the parameters in Ref.~\cite{KOB95}
yield $\chi^2/N_{d.f.} = 7.5 $. 
This parameterization, in contrast with the two other ones presented in 
this paper, can
be extrapolated well above 7 fm$^{-1}$, albeit with some theoretical
prejudice. We confirm the observation of Refs.~\cite{KOB95,GAR99} 
that the double helicity flip
transition amplitude $g_2$ has a magnitude comparable to the zero helicity flip
amplitude $g_0$ in the $Q$-range considered here, which means that these
amplitudes are not in the asymptotic regime expected from pQCD.

{\bf 3.3 Parameterization III}

The third parameterization (III) employs a Sum-of-Gaussians (SOG) \cite{SIC74}.
The form factors are written as
\begin{equation} \label{eq:sog}
G_X(Q) = G_X(0) \cdot e^{-\frac{1}{4}Q^{2}\gamma^{2}} 
\sum _{i=1}^{25} \frac{A_{i}}{1 + 2
R_{i}^{2}/\gamma ^{2}} \left(\cos (QR_{i}) + \frac{2 R_{i}^{2}}{\gamma ^{2}}
\frac{\sin (QR_{i})}{QR_{i}}\right )
\end{equation}
Although our interest here lies in its $Q$-space version, the parameterization
is better described in configuration space where it corresponds to a 
density $\rho(R)$  written as a sum of Gaussians placed at arbitrary 
radii $R_{i}$, with amplitudes $A_{i}$  fitted to the data, and a fixed 
width $\gamma$. The distance $R$ refers to the distance of the nucleons 
to the deuteron center of mass. The parameterization 
represents a totally general basis and the following applied restrictions are 
justified on physics grounds. First, 
one does not expect structures smaller than the size of the
nucleon, which determines the width $\gamma$ to be the size of the
proton ($\gamma \sqrt{3/2} = 0.8$ fm). 
Second, the spacing between Gaussians is chosen slightly smaller than this width: 
0.4 fm or 0.5 fm.
Third, the Gaussians are placed 
at radii $R_{i} \leq R_{max} = 10$ fm, which is justified given 
the fact that one can easily specify the radius at which the 
tails of densities give no significant $(< 10^{-3})$ contribution 
to $G_X(Q)$. In addition, outside the range of the NN--force, 
the deuteron wave functions have an analytic form which is well 
known and depends only on the deuteron binding energy. 
Thus, for radii $R_{i} \geq 4$ fm, one can impose this shape and 
fix the ratio of the amplitudes $A_i$.
Each form factor is then determined with 11 free parameters: 10 Gaussian
amplitudes $A_1$ to $A_{10}$, corresponding to $R_i < 4$ fm, and one overall 
amplitude for the shape-given tail at $R \geq 4$ fm.
With a total of 33 independent parameters,
a $\chi^2/N_{d.f.}$ of 1.5 is obtained in the fit.

{\bf 3.4 Results and discussion}

The resulting form factors from the three parameterizations are 
shown in Fig. 1. As functions of two variables ($Q$ and $\theta_e$),
the fitted quantities cannot be easily represented together 
with the parameterizations.
In order to illustrate the quality of the fits, we present plots of
relative differences of $A$ and $B$,
and of $\tilde{t}_{20}(Q)$
in Fig.~2. $t_{21}$ and $t_{22}$ are equally well fitted, which constitutes,
within experimental uncertainties, an indication of the coherence of 
equations (\ref{eq:sigma},\ref{eq:t20},\ref{eq:t21},\ref{eq:t22}),
and therefore of the consistency of the one-photon exchange approximation. 

From the average and dispersion between the three parameterizations,
combined with the fit uncertainty on $Q_C^0$, 
the node of the charge form factor is determined to be located at
$4.21 \pm 0.08$ fm$^{-1}$, a value governed by the $t_{20}$ results of
Refs.~\cite{ABB00,GAR94}. Assuming as we do here implicitly that
these two data sets have the same weight,
the location of this node is not quite consistent with 
a relation between the two- and three-nucleon isoscalar charge form factors,
established with various $N-N$ potentials~\cite{HEN95}.
The secondary maximum of $|G_C|$ is very flat, so that its location 
($5.3 \pm .5$ fm$^{-1}$) is not determined very precisely. 
Its magnitude ($.0038 \pm .0003$) is clearly inconsistent with the
corresponding one 
of the three-nucleon isoscalar charge form factor, still within the
same model calculations~\cite{HEN95}.
The $t_{21}$ results of Ref.~\cite{ABB00},
though of limited accuracy, help confirm a node of the magnetic
form factor \cite{BOS90} at $7.2 \pm 0.3$ fm$^{-1}$. 
As for the first node of $G_Q$, 
according to most theoretical models, 
it should appear at a higher value of $Q$, above the
range where our parameterization method applies.
The value $Q_Q^0=7.7 \pm 0.6$  fm$^{-1}$ given by parameterization I
is probably the smallest possible value allowed by the present data.
It is due to this parameterization following the downward trend
of the $t_{20}$ data point at the highest $Q$ (see Fig.~2). This 
trend however is not statistically significant.
Parameterization II,
when extrapolated, suggests a much higher value of $Q$ for the node of $G_Q$.  
Finally, from 
\begin{equation}
r^2 \equiv \left. - 6\ {dG_C \over dQ^2}\right|_{Q^2=0} = 6\left[ {a_C}_1 + (Q_C^0)^{-2}\right],
\end{equation}
we calculate
the root mean square charge radius of the deuteron 
$r=2.094\pm 0.003$ (stat.) $\pm 0.009$ (syst.) fm.
The statistical uncertainty is given by the error matrix from parameterization I, while
the systematic uncertainty is evaluated with parameterization III (see above remark
about normalization uncertainties on individual data sets).
This radius is 1.7\% smaller than the 
value $r=2.130$ fm reported in \cite{SIC98}, consistent with expectations
in the absence of  corrections due to Coulomb distortion.

\vspace{5mm}
{\bf 4 Conclusion}

The extraction of the charge form factors $G_C$ and $G_Q$ 
from experiment,
at given values of $Q$, has been reexamined.
The solutions were expressed in the most compact and physical way,
while a new treatment of errors was applied to polarization data
at or beyond  physical limits.
The existing electron-deuteron elastic scattering data were used
for direct parameterizations of the three deuteron electromagnetic form
factors, up to $Q=7$ fm$^{-1}$. 
The numerical results may be requested from the 
authors\footnote{Contacts: 
		jball@cea.fr (parameterizations I and II), 
		jourdan@ubaclu.unibas.ch (III).
		}
and will be updated as new data become available in the future.
The inferred value of $Q_{min}\simeq 3.3$ fm$^{-1}$ corresponding to the minimum of
$\tilde{t}_{20}$ could be used, or recalculated with such global fits,
for future experiments in this $Q$-range~\cite{TUR98,VEPP}, in order to resolve 
the discussed ambiguities in the form factors calculation.
These future experiments should help confirm, or adjust, the
exact value of the node of the charge form factor: this location is
sensitive to the strength of the $N-N$ repulsive core, to the size of the
isoscalar meson exchange contributions and to relativistic corrections. 
The observation of the node of the magnetic form factor~\cite{BOS90,ABB00}
should be confirmed in a more precise experiment~\cite{MAKIS}.
Together with the determination of the
secondary maximum of $|G_C|$~\cite{ABB00}, this would complete the full
characterization of the deuteron electromagnetic structure up to $Q\simeq 7$ fm$^{-1}$.

\acknowledgements{
	The authors gratefully acknowledge discussions with A. Kobushkin and I. Sick.

	This work was supported by the French Centre National de la
Recherche Scientifique and Commissariat \`a l'Energie Atomique, the
Swiss National Science Foundation, the U.S. Department of Energy and
National Science Foundation, and the K.C. Wong Foundation.
	}


\begin{table*}
\begin{tabular}{cccccc}
\vspace*{-3mm} 
$Q$ & 
$t_{20}(70^{\circ})$ & $\tilde{t}_{20}$ & $G_C$ & $G_Q$ & Ref.\\ 
(fm$^{-1}$)  &  
&   &   & 	\\ \hline
\small
0.86   	
	& -.30 \footnotesize  ($\pm$.16) 		  & -.30 \footnotesize  ($\pm$.16)  
	& .627 \footnotesize  ($\pm$.011)  		  & 47. \footnotesize  ($\pm$25.) 	
	& \cite{DMI85} 	\\
1.15	
	& -.181 \footnotesize  ($\pm$.070) 		  & -.178 \footnotesize  ($\pm$.071)
	& .474 \footnotesize  ($\pm$.008)  		  & 12.0 \footnotesize  ($\pm$4.7)
	& \cite{VOI86}	\\
1.58	
	& -.400 \footnotesize  ($\pm$.037)                & -.402 \footnotesize  ($\pm$.038) 
	& .289 \footnotesize  ($\pm$.006)                 & 8.66 \footnotesize  ($\pm$.81)
	& \cite{FER96}	\\
1.74	
	& -.420 \footnotesize  ($\pm$.060)                & -.423 \footnotesize  ($\pm$.063)
	& .238 \footnotesize  ($\pm$.005)                 & 6.19 \footnotesize  ($\pm$.90)
	& \cite{SCH84}	\\
2.026	
	& -.713 \footnotesize  ($\pm$.090)                & -.734 \footnotesize  ($\pm$.095)
	& .160 \footnotesize  ($\pm$.005)                 & 5.51 \footnotesize  ($\pm$.73)
	& \cite{BOU99}	\\ 
2.03	
	& -.590 \footnotesize  ($\pm$.130)                & -.604 \footnotesize  ($\pm$.138)
	& .163 \footnotesize  ($\pm$.005)                 & 4.50 \footnotesize  ($\pm$1.02)
	& \cite{SCH84}	\\
2.352	
	& -.896 \footnotesize  ($\pm$.093)                & -.945 \footnotesize  ($\pm$.101)
	& .100 \footnotesize  ($\pm$.004)                 & 3.49 \footnotesize  ($\pm$.41)
	& \cite{BOU99}	\\
2.49	
	& -.751 \footnotesize  ($\pm$.153)                & -.792 \footnotesize  ($\pm$.169)
	& .087 \footnotesize  ($\pm$.004)                 & 2.17 \footnotesize  ($\pm$.48)
	& \cite{GIL90}	\\
2.788	
	& -1.334 \footnotesize  ($\pm$.233)               & -1.473 \footnotesize  ($\pm$.267)
	&  3.71 \footnotesize  ($^{+1.47}_{-0.11}$)$\times$10$^{-2}$ & 2.59 \footnotesize  ($\pm$.073)
	& \cite{BOU99}	\\
2.93	
	& -1.255 \footnotesize  ($\pm$.299)               & -1.401 \footnotesize  ($\pm$.347)
	& 3.45 \footnotesize  ($^{+1.22}_{-0.39}$)$\times$10$^{-2}$ & 1.85 \footnotesize  ($^{+.12}_{-.64}$)
	& \cite{GIL90}	\\
3.566	
	& -1.87 \footnotesize  ($\pm$1.04)                & -2.20 \footnotesize  ($\pm$1.26)
	& 1.53 \footnotesize  ($^{+0.06}_{-1.38}$)$\times$10$^{-2}$ & .651 \footnotesize  ($^{+.147}_{-.023}$)
	& \cite{BOD91}	\\
3.78	
	& -1.278 \footnotesize  ($\pm$.186)               & -1.476 \footnotesize  ($\pm$.228)
	& 1.25 \footnotesize  ($^{+.05}_{-.55}$)$\times$10$^{-2}$ & .474 \footnotesize  ($^{+.078}_{-.018}$)
	& \cite{GAR94}	\\
4.09	
	& -.534 \footnotesize  ($\pm$.163)                & -.567 \footnotesize  ($\pm$.193)
	& -1.14 \footnotesize  ($\pm$1.6)$\times$10$^{-3}$ & .383 \footnotesize  ($\pm$.015)
	& \cite{ABB00}	\\
4.22	
	& -.833 \footnotesize  ($\pm$.153)                & -.913 \footnotesize  ($\pm$.179)
	& 1.63 \footnotesize  ($^{+1.61}_{-1.44}$)$\times$10$^{-3}$ & .325 \footnotesize  ($\pm$.013)
	& \cite{GAR94}	\\
4.46	
	& -.324 \footnotesize  ($\pm$.089)                & -.320 \footnotesize  ($\pm$.100)
	& -2.39 \footnotesize  ($\pm$.61)$\times$10$^{-3}$ & .245 \footnotesize  ($\pm$.010)
	& \cite{ABB00}	\\
4.62	
	& -.411 \footnotesize  ($\pm$.187)                & -.417 \footnotesize  ($\pm$.207)
	& -1.63 \footnotesize  ($\pm$1.14)$\times$10$^{-3}$ & .208 \footnotesize  ($\pm$.009)
	& \cite{GAR94}	\\
5.09	
	& .178 \footnotesize  ($\pm$.053)                 & .208 \footnotesize  $\pm$(.056)
	& -3.87 \footnotesize  ($\pm$0.30)$\times$10$^{-3}$ & .119 \footnotesize  ($\pm$.006)
	& \cite{ABB00}	\\
5.47	
	& .292 \footnotesize  ($\pm$.073)                 & .312 \footnotesize  ($\pm$.075)
	& -3.48 \footnotesize  ($\pm$0.32)$\times$10$^{-3}$ & .080 \footnotesize  ($\pm$.004)
	& \cite{ABB00}	\\
	\vspace*{-2mm}
6.15	
	& .621 \footnotesize  ($\pm$.168)                 & .630 \footnotesize  ($\pm$.170)
	& -3.19 \footnotesize  ($\pm$0.55)$\times$10$^{-3}$ & .034 \footnotesize  ($^{+.005}_{-.006}$)
	&   		\\	
	& & 
	& -4.20 \footnotesize  ($^{+.42}_{-.32}$)$\times$10$^{-3}$& .019 \footnotesize  ($\pm$.007)
	& \cite{ABB00}	\\
	\vspace*{-2mm}
6.64    
	& .476 \footnotesize  ($\pm$.189)                 & .478 \footnotesize  ($\pm$.189)  
	& -1.89 \footnotesize  ($\pm$0.38)$\times$10$^{-3}$ & .023 \footnotesize  ($^{+.002}_{-.003}$)	
	&   		\\
	\vspace*{-5mm}
	& & 
	& -3.13 \footnotesize  ($^{+.24}_{-.19}$)$\times$10$^{-3}$& .008 \footnotesize  ($\pm$.004)
	&  \cite{ABB00}	\\  
\normalsize
\end{tabular}
\normalsize
\caption{Calculated values of $t_{20}(70^{\circ})$, $\tilde{t}_{20}$, $G_C$ and
		$G_Q$ corresponding to all $t_{20}$ measurements. 
	In parantheses, statistical and systematic uncertainties are 
	added in quadrature.
	For the last two points, the two solutions are given (see text).
	}
\label{tab2}
\end{table*}
\begin{figure}
\begin{caption}{ }
Deuteron form factors $G_C$, $G_Q$ and $G_M$ as a function of $Q$. 
The data for $G_C$ and $G_Q$ are from Table 1,
corresponding to $t_{20}$ measurements of Refs
\cite{ABB00} (solid diamonds, and open diamonds for the second solution),
		 \cite{BOD91} (star),
		 \cite{BOU99} (open squares),
		 \cite{DMI85,VOI86} (triangles up),
		 \cite{FER96} (open circle), 
		 \cite{GAR94} (full squares), 
		 \cite{GIL90} (triangles down), 
		 \cite{SCH84} (full circles).
		The  $G_M$ data corresponds to the $B$ measurements of Refs.
		 \cite{AUF85} (open diamonds),
		 \cite{BOS90} (open circles), 
		 \cite{CRA85} (stars), 
		 \cite{SIM81} (full circles).
		 The curves are from our parameterizations
		  I (solid line), II (dot-dashed line) and III (short dashed line).
\end{caption}

\end{figure}

\begin{figure}[h]
\begin{caption}{ }
		(a) $\Delta A/A$, in \%: deviation of $A$ with respect to parameterization I,
		arbitrarily taken as a reference line; for clarity only the data from 
		Refs~\cite{ABB99} (full diamonds), 
		\cite{ALE99} (full circles),
		\cite{ARN75} (open circles), 
		\cite{ELI69} (triangles),
		\cite{PLA90} (open diamonds)
		 are reported.
   		(b) $\Delta B/B$, in \%. 
		(c) $\tilde{t}_{20}$, with physical domain delimited by dotted lines. 
		For $B$ and $\tilde{t}_{20}$ data legend, as well as curves legend, 
		see Fig. 1.	
\end{caption}

\end{figure}
\newpage
\begin{figure}
\begin{center}
\mbox{ \epsfxsize=15.cm\leavevmode \epsffile{fig1.epsi}}
\end{center}
\end{figure}

\newpage
\begin{figure}
\begin{center}
\mbox{ \epsfxsize=15.cm\leavevmode\epsffile{fig2.epsi}}
\end{center}
\end{figure}
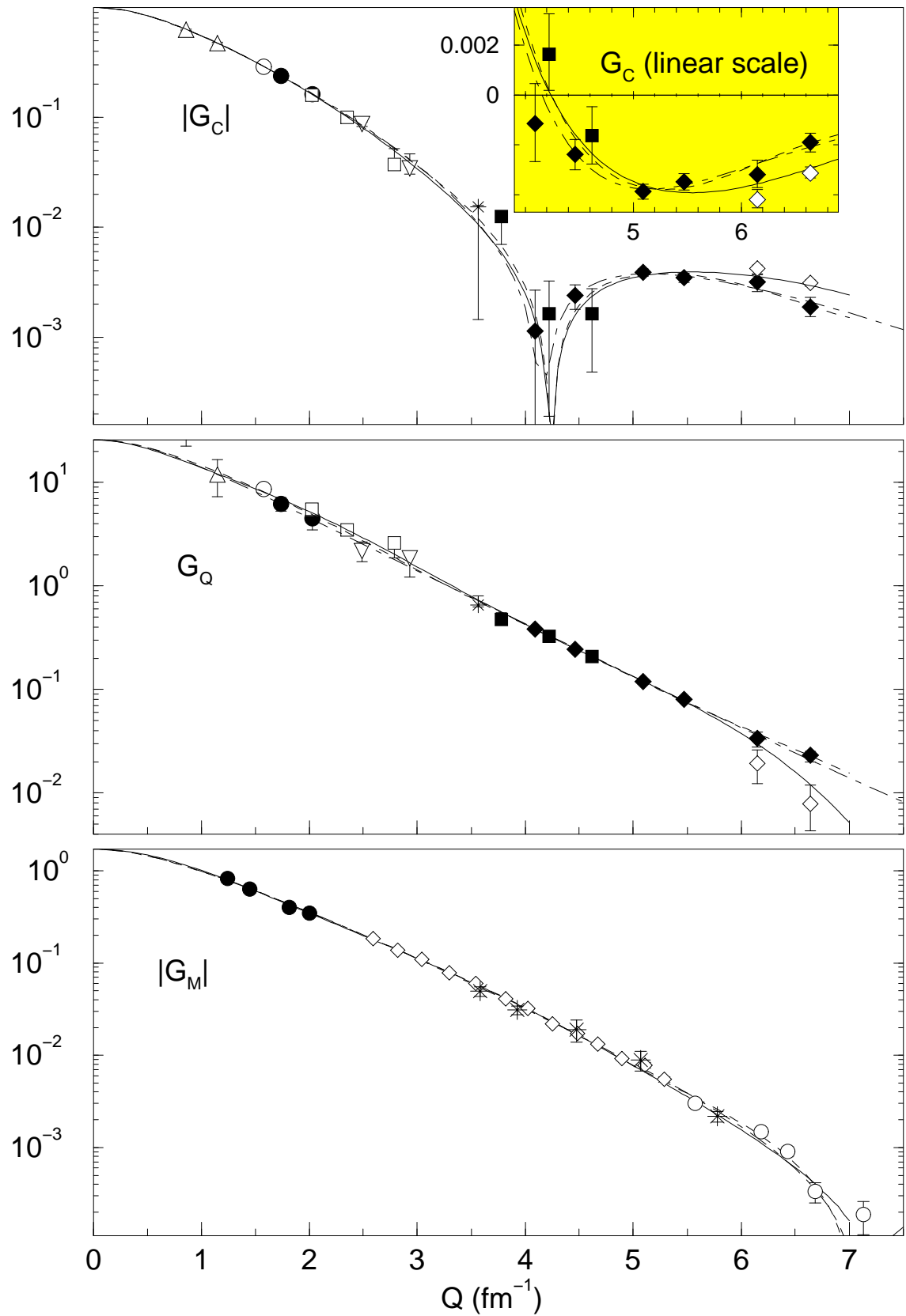
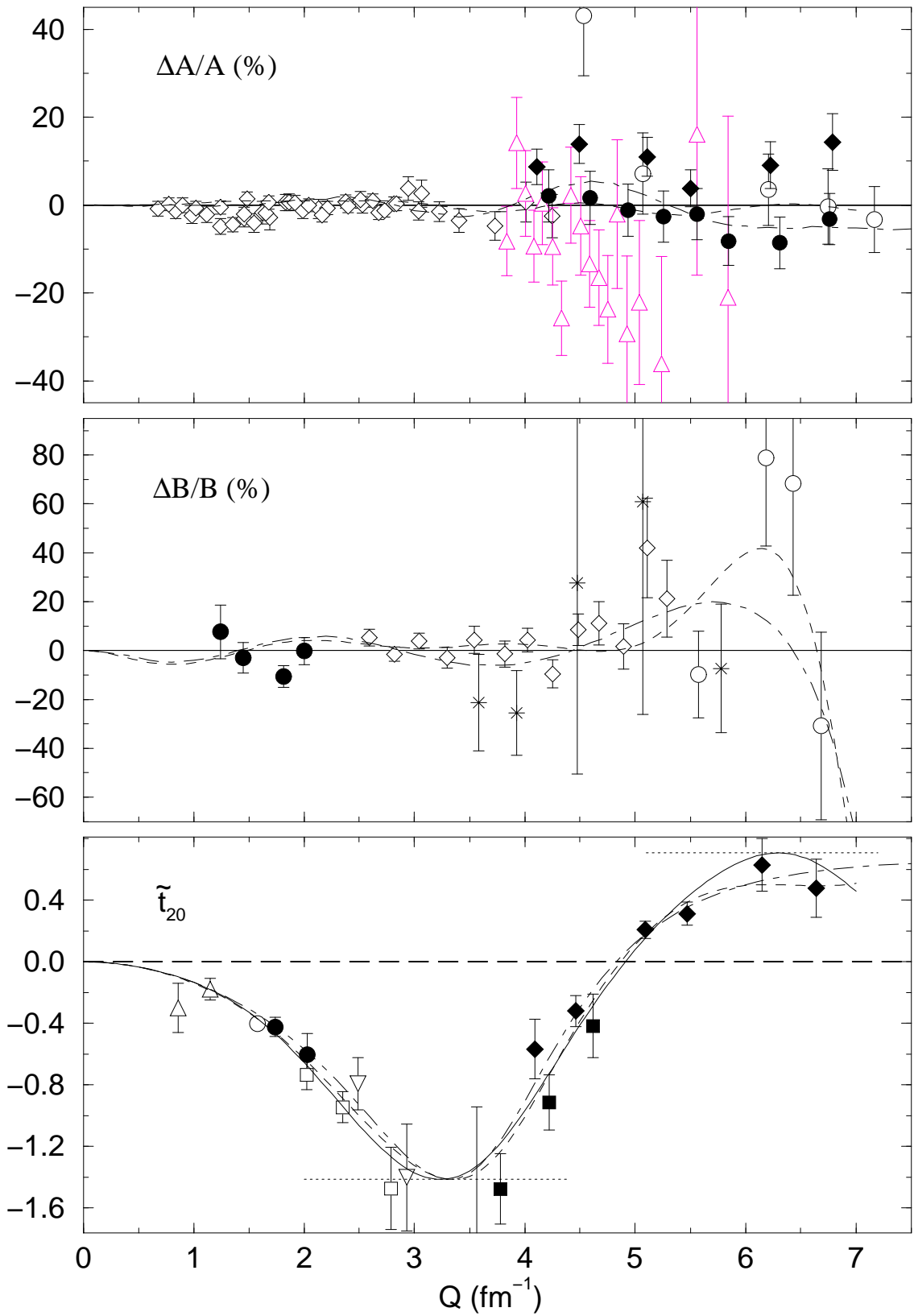
\end{document}